\documentclass[
reprint,
unsortedaddress,
nofootinbib,
nobibnotes,
amsmath,amssymb,
aps,
prd,
superscriptaddress,
]{revtex4-1}

\usepackage{graphicx}% Include figure files
\usepackage{dcolumn}% Align table columns on decimal point
\usepackage{bm}% bold math
\usepackage{url	}
\usepackage{float}

\usepackage{hyperref}% add hypertext capabilities
\usepackage{cleveref}
\usepackage[dvipsnames]{xcolor}
\usepackage[normalem]{ulem}

\newcommand*{\ksM}{\text{km/s Mpc$^{-1} $}}

\hypersetup{
    colorlinks,
    citecolor=[rgb]{0.16,0.67,0.53},
    linkcolor=[rgb]{.25,.41,.88},
}

\newcommand*\aap{Astron.~Astrophys.}

\newcommand*\apss{Ap\&SS}

\newcommand*\jcap{J. Cosmology Astropart. Phys.}

\newcommand*\mnras{Mon.~Not.~Roy.~Astron.~Soc.}

\newcommand*\pasp{PASP}

\newcommand{\diff}{\mathrm{d}}
\defcitealias{Riess18_H0}{R18}
\defcitealias{Vattis19}{V19}
\defcitealias{Verde17}{V17}

\begin{document}

%\preprint{APS/123-QED}
% Seeking sources of $H_0$-tension in Cosmic Microwave Background data
\title{Sources of $H_0$-tensions in dark energy scenarios}% Force line breaks with \\

\author{Balakrishna S. Haridasu}
\email{haridasu@roma2.infn.it}

\affiliation{Dipartimento di Fisica, Universit\`a di Roma "Tor Vergata", Via della Ricerca Scientifica 1, I-00133, Roma, Italy}
\affiliation{Sezione INFN, Universit\`a di Roma "Tor Vergata", Via della Ricerca Scientifica 1, I-00133, Roma, Italy}

\author{Matteo Viel}
\email{viel@sissa.it}
\affiliation{SISSA-International School for Advanced Studies, Via Bonomea 265, 34136 Trieste, Italy}
\affiliation{INFN, Sezione di Trieste, Via Valerio 2, I-34127 Trieste, Italy}
\affiliation{INAF - Osservatorio Astronomico di Trieste, Via G. B. Tiepolo 11, I-34143 Trieste, Italy}
\affiliation{IFPU, Institute for Fundamental Physics of the Universe, via Beirut 2, 34151 Trieste, Italy}

\author{Nicola Vittorio}
\email{nicola.vittorio@roma2.infn.it}

\affiliation{Dipartimento di Fisica, Universit\`a di Roma "Tor Vergata", Via della Ricerca Scientifica 1, I-00133, Roma, Italy}
\affiliation{Sezione INFN, Universit\`a di Roma "Tor Vergata", Via della Ricerca Scientifica 1, I-00133, Roma, Italy}
\date{\today}

\begin{abstract}
    By focusing on the simple $w\neq-1$ extension to $\Lambda$CDM, we assess which epoch(s) possibly source the $H_0$-tension. We consider Cosmic Microwave Background (CMB) data in three possible ways: $i)$ complete CMB data; $ii)$ excluding the $l<30$ temperature and polarization likelihoods; $iii)$ imposing early universe priors, that disentangle early and late time physics. Through a joint analysis with low-redshift supernovae type-Ia and gravitationally lensed time delay datasets, {and neglecting galaxy clustering Baryonic Acoustic Oscillation (BAO) data}, we find that the inclusion of early universe CMB priors is consistent with the local estimate of $H_0$ while excluding the low-$l$+lowE likelihoods mildly relaxes the tension. This is in contrast to joint analyses with the complete CMB data. Our simple implementation of contrasting the effect of different CMB priors on the $H_0$ estimate shows that the early universe information from the CMB data when decoupled from late-times physics could be in agreement with a higher value of $H_0$. {We also find no evidence for the early dark energy model using only the early universe physics within the CMB data. Finally using the BAO data in different redshift ranges to perform inverse distance ladder analysis, we find that the early universe modifications, while being perfectly capable of alleviating the $H_0$-tension when including the BAO galaxy clustering data, would be at odds with the Ly-$\alpha$ BAO data due to the difference in $r_{\rm d}\, vs.\, H_0$ correlation between the two BAO datasets.} We therefore infer and speculate that source for the $H_0$-tension between CMB and local estimates could possibly originate in the modeling of late-time physics within the CMB analysis. This in turn recasts the $H_0$-tension as an effect of late-time physics in CMB, instead of the current early-time CMB vs. local late-time physics perspective.
\end{abstract}

\maketitle

%\tableofcontents

\section{Introduction}
The $H_0$-tension is now significant at confidence levels $\gtrsim 5\sigma$  \cite{Wong19, Riess19_Nat, Verde:2019ivm}, making it the pressing issue to be resolved within the current/near-future cosmological scenario and possibly getting evidences for new physics beyond the standard $\Lambda$CDM model. Several proposals have been put forward attempting to resolve this tension, which can be classified based on the epoch(s) at which the modifications are suggested: high-redshift pre-recombination (early-time), low-redshift (late-time) and the local resolution. The early-time modifications \cite{Khosravi17, Banihashemi18} such as, early dark energy \cite{Poulin18, Karwal16, xiaviel09, Ye20, Ye20b,  Mortsell18, Das20, Sakstein19, Khoraminezhad20}, interacting dark energy \cite{Raveri18, Ko16, Archidiacono19} or early modified gravity \citep{Braglia20}, among others \cite{smith19, Lin19, Agrawal19, Lin20} intend to modify the inferred scale of the sound horizon keeping the angular scales constrained by the Cosmic Microwave Background (CMB) data \cite{Planck18_parameters} unchanged. {Late-time modifications such as the decaying dark matter models \cite{Blackadder14, Vattis19, Hryczuk20, Choi19, Poulin16}, which aim to increase the expansion rate at low-redshifts and hence the $H_0$, are also seemingly unable to relieve the tension \cite{Haridasu20, Abellan20, Clark20}} (see also \cite{ Raveri19}). Several attempts in to late-time modifications of the background expansion history seem inadequate to relieve the tension \cite{Verde16, Poulin18a, Yang18a, Aylor18, Lemos18, Arendse19, Lyu20}. On the other hand, the local resolutions \cite{Keenan13, Shanks19} rely on correctly assessing the local `inhomogeneous' distribution of matter, essentially modifying the local estimate \cite{Hoscheit17, Shanks19, Schoneberg19} to be in agreement with the lower high-redshift $\Lambda$CDM based $H_0$. However, this approach is in contrast with the dynamics constrained by the Supernovae datasets \cite{Kenworthy19, Lukovic19}, which do no allow sufficient variation in the local value necessary to completely resolve the tension. And would also remain at a disadvantage to explain the higher values of $H_0$ obtained by the Gravitational lensing time delay datasets \cite{Wong19, Birrer19, Millon19, Birrer20}. 

In this work we investigate the inferred $H_0$ value from the different components of CMB data, through selectively contrasting the CMB likelihoods with a combination of low-redshift data, using the simple one parameter extension allowing dark energy equation of state (EoS) $w \neq -1$ ($w$CDM). This extension was explored as a resolution to $H_0$-tension where the EoS is required to be phantom-like ($w<-1$) \cite{DiValentino17, Vagnozzi19, Alestas20}, however with some non-standard statistical interpretations. This model is clearly a low-redshift modification, as the dark energy component affects only the late-time evolution. By construction, this also allows us to make inferences for the $w = -1$, $\Lambda$CDM model. In the current work, we rely on  a $w$CDM  and an early dark energy model, owing to the fact that the $\Omega_{\rm k} \neq 0$, extension is well-known to be in even stronger discrepancy with the high values of local estimate, having $H_0 \sim 55 \, \ksM$ \cite{Planck18_parameters} and an even further discrepancy with Baryon Acoustic Oscillations data, pointed out recently in \citep{Handley19, DiValentino19}. However, as we discuss later in \Cref{sec:Results}, our final inference would not depend on this choice.

We use the low-redshift Supernovae Type Ia (SN) \cite{Scolnic17} dataset which capture very well the dynamics of the background expansion and the Gravitational lensing time delay (SL) \cite{Birrer19, Wong19} dataset, that complements the SN by providing constraints on $H_0$, which is also in agreement with the local estimate. Incidentally, there is a good agreement in the $H_0$ vs. $w$ parameter space constraints obtained form the low-redshift SL \cite{Taubenberger19, Wong19} and high-redshift Cosmic Microwave Background (CMB) \cite{Planck18_parameters} datasets, which is also a  reason for the analyses performed here. It is also well-known that the CMB based constraints on $w$ are phantom-like at $\sim 95\%$ C.L. and in tentative tentative tension with $\Lambda$CDM model. In principle one can find a three-way agreement between local model-independent SN based $H_0 = 73.48\pm 1.42 \,\ksM$ in \citet{Riess18_H0} (hereafter \citetalias{Riess18_H0}), high-redshift CMB based $H_0>70.4\, \ksM (95\% {\rm C.L.})$ \cite{Planck18_parameters} and low-redshift SL based $H_0 = 80.8^{+5.3}_{-7.1} \ksM$ \cite{Taubenberger19} for a $w<-1$ dynamic dark energy extension. 

Traditionally, CMB based priors (reduced likelihood) have been through the distance, angular scales at the last scattering surface, the so called shift parameters \cite{Wang06, Wang07}. This implementation however remains model-dependent for the extrapolated low-redshift behavior and need to be assessed separately for distinct cosmologies \cite{Komatsu08, Planck15_DE, Chen18} (see \citet{Zhai19} for a more recent extended discussion). In this context, we firstly conduct a `selective' contrasting analysis using combination of high-$l$ ($l\geq30$) and CMB-$lensing$ likelihoods from latest \textit{Planck} dataset, which is expected to minimize the effects of late-time physics \cite{Verde17}. Following which, we also use constraints obtained from this analysis as alternate priors, as proposed in Ref.~\citep{Verde17} (earlier suggested in \cite{Vonlanthen10, Audren12, Audren14}), which intend to disentangle the late-time effects and obtain early-time priors from the CMB data which are independent of standard model-extensions such as the $w \neq -1$ and $\Omega_{\rm k} \neq 0$, that affect the low-redshift evolution. Clearly, exclusion of data is not intended as a resolution of the $H_0$-tension, however the goal is to assess which epoch(s) within the CMB data gives rise to the tension with the local estimate and is in turn expected to possibly point towards desirable extensions/modifications of the current cosmological scenario. {We also perform this analysis for the early dark energy model \cite{Poulin18, Ivanov20, Hill20}, which cannot however be interpreted as early universe priors due to the specificity of the assumed model, but will help to assess the evidence for the model, when utilizing only the early-time information in the CMB data. }

In the main analysis, we do {\it not} include Baryon Acoustic Oscillations (BAO) \cite{Eisenstein05, Alam16, Bautista17, Bourboux17} datasets at low-redshifts, as possible variations in the angular scales at high-redshift might be overlooked, which are well constrained by the CMB datasets and are necessarily assumed as fiducial cosmology, when obtaining the BAO observables. It is also well-known that the BAO constraints on the dark energy EoS ($w$) are not in immediate agreement with the CMB based phantom-like inferences \cite{Aubourg14, Haridasu17_bao, Bernal16, Park19}. However, see also relevant analyses made: i) Ref.~\citep{Anselmi18} (see also \cite{Anselmi17, Anselmi17b}), which claims an increased uncertainty of the BAO observables on the account of assumed fiducial model through a purely geometric formalism; ii) Ref.~\citep{Ivanov19}, where not assuming a CMB based fiducial cosmology shows increased uncertainty for varying dark energy EoS model (see Fig. 5 therein). These analyses suggest that the BAO observables might support a $w<-1$ scenario or a variation in the constraints of $r_d \times H_0$ combination, through increased uncertainties in the current observables. In effect, the mild disagreements between the constraints obtained from the isotropic and anisotropic components of BAO data reported in \cite{Haridasu17_bao}, might be pointing out for either underestimated uncertainty or an induced bias due to the assumed fiducial cosmology; all these efforts would require further investigation. {However, we do rely on the BAO data to perform a simple inverse distance ladder analysis by assuming $\Omega_{\rm b}h^2$ priors from the CMB data. Which aids the discussion regarding the sufficiency of an early universe modification to alleviate the $H_0$-tension. Recently \citet{Hill20, Ivanov20} (see also \cite{Lucca20}) have argued that the early dark energy model is not adequate to alleviate the tension owing to no evidence with the CMB data alone and discrepancy with Large Scale Structure (LSS) data \cite{Abbott17, Hildebrandt16, Hildebrandt18, Hikage18}, respectively. However, see also \cite{Smith20, Niedermann20, Chudaykin20b} where the LSS data is instead shown to be agreement with the EDE model. }

The current paper is organized as follows: In \Cref{sec:Theory1} we describe the theoretical models implemented, followed by the description of data in \Cref{sec:Data}. The results and discussions are presented in \Cref{sec:Results} and finally we conclude in \Cref{sec:Conclusions}.

\section{Theory and Modeling}
\label{sec:Theory1}
We test the flat ($\Omega_{\rm k} = 0$) dynamical dark energy extension, with constant EoS $w \neq -1$, for which the expansion rate is written as,

\begin{equation}\label{eq:Hubble}
\frac{H(a)^2}{H_0^2} \equiv  E(a)^2 = \left[ \frac{\Omega_{r}}{a^4} + \frac{\Omega_{m}}{a^3} + \Omega_{de} f(a)  \right], 
\end{equation}

where $H(a)$ is the Hubble rate in terms of the scale factor $a$, $\Omega_i = \rho_{i0}/\rho_{c0}$ is the current energy density of the $i^{th}$ component normalized to the current critical density of the Universe $\rho_{c0} = 3 H^2_0 m^2_{pl}$ and subscript `0' corresponding to the quantities measured today (a = 1). The time dependence of the dark energy component is modeled through the the function $f(a)$ which can be written from continuity equation as,
\begin{equation}\label{eq:f(a)}
f(a) = \exp{\left[ 3 \int_{a}^{1} \frac{d \tilde{a}}{\tilde{a}} (1+w_{\phi}(\tilde{a}))\right]}.
\end{equation}
which accounts to $a^{-3(1+w)}$ for a constant EoS, $w$CDM model. Distances are as usual estimated in the standard way for a flat background through transverse comoving distance $D_{\rm M}(z) = \frac{c}{H_{0}}\int_{0}^{z} \frac{\diff\xi}{E(\xi)}$. Throughout the analysis we fix the radiation density\footnote{We assume the total radiation energy density as $\Omega_{r0} = 4.18343\times 10^{-5}$, corresponding to the present $T_{\rm CMB} = 2.7255\, \rm{K}$ \citep{Fixsen09}, which is the same implementation in \textit{Planck} 2018 \cite{Planck18_parameters} analysis.} based on the usual implementation as in \textit{Planck} 2018 analysis \cite{Planck18_parameters}. 

{Alongside the $w$CDM model we also perform minimal analysis for the Early Dark Energy (EDE/3pEDE) model\footnote{We use the publicly available EDE modification of \texttt{CLASS}, \texttt{CLASS\_EDE} provided by \citet{Hill20} at \href{https://github.com/mwt5345/class_ede}{https://github.com/mwt5345/class\_ede}.}. In this particular model a light scalar field is conjectured which allows the effective cosmological constant to dynamically decay. The potential of the scalar field $\phi$ can be written as,}

\begin{equation}
    V(\phi) = m^2f^2(1-cos(\phi/f))^n. 
\end{equation}
{The physics of the scalar field can be described through effective field parameters $z_{c}$, $f_{\rm EDE} = \rho_{\rm EDE}(z_{\rm c})/3m_{\rm pl}^2H(z_{\rm c})^2 $ and $\theta_{\rm i} = \phi_{\rm i}/f$. Here $z_{\rm c}$ denotes the redshift at which the EDE contributes the most to the total energy density. Here we have kept the theoretical description of the EDE model to a minimum, please refer to \citet{Poulin18, Hill20,Ivanov20} for an elaborate description of the theory and the modeling. Therefore, the effective dynamics of the scalar field will be described by three additional parameters $\Theta_{\rm EDE} \equiv \{f_{\rm EDE}, \log_{10}(z_{\rm c}), \theta_{\rm i}\}$.}

The transverse comoving distance $D_{\rm M}$ is related to angular diameter distance $D_{\rm A}(z) = D_{\rm M}(z)/(1+z)$, which is used in the construction of the shift parameters:
\begin{align}
\label{eq:Distance_Priors}
    R(z_{*}) &= \frac{(1+z_{*})D_{\rm A}(z_{*})\sqrt{\Omega_m H_0^2}}{c} \\
    l_{\rm A}(z_{*}) &= \frac{\pi (1+z_{*})D_{\rm A}(z_{*})}{r_s(z_{*})}
\end{align}
where $z_{*}$ is redshift to the photon decoupling epoch and $r_s$ is the sound horizon. Also, when appropriate we use the numerical fitting formula for the sound horizon at drag epoch $r_{\rm d}$,  provided in \citet{Aubourg14}.

\section{Data}
\label{sec:Data}

\subsection{Low-redshift probes}
For low-redshift data we consider Supernovae Type Ia and Strong lenses time delays.

\textit{Supernovae Type Ia} (SN): We use the Pantheon compilation of $\sim 1050$ supernovae (SNe) observations presented in \cite{Scolnic17} have improved the statistical precision and the highest redshift ($z\sim 2$) to which the distances are measured. The likelihood for the SN dataset is implemented as suggested in the release \cite{Scolnic17}. 

\textit{Strong lenses} (SL): A combination of six gravitationally lensed time delay systems as were implemented earlier in \cite{Wong19, Lyu20} is considered. We follow the same procedure implemented in \citet{Wong19} closely replicating their results also in combination with the CMB datasets. The dataset implemented in \citet{Wong19}, is an improvement over the 4 lenses dataset implemented earlier in \citet{Taubenberger19}, with a corresponding improvement in the constraints on $H_0$, reaching a $2.4\%$ precision.

{\textit{Baryon Acoustic Oscillations} (BAO): We use a compilation of angular diameter distance $D_{\rm A}(z)/r_d$ and the Hubble rate $H(z)\times r_d$, at low-redshift taken from \cite{Alam16} (DR12), at intermediate redshifts from \cite{Zhao18_dr14} (DR14) and finally the high-redshift measurements of Ly-$\alpha$ forest, auto-correlation and the Ly-$\alpha$ and quasars cross-correlation taken from \cite{Blomqvist19} and \cite{SainteAgathe19} (Ly$\alpha$19), respectively.}

\subsection{High-redshift likelihood(s)}
\label{subsec:CMB}

\textit{Early Universe priors} (EUp-15/EUp-18): Our goal is to disentangle the early-time constraints from the effects of the low-redshift behavior of models, as in \cite{Verde17} (hereafter \citetalias{Verde17}). The constraints on the energy densities and expansion rate at the recombination era have been reported and shown to be independent of an additional degree of freedom, in particular $w, \Omega_{\rm k}$ (c.f. Table 2 therein). This is clearly evident of the well-expected fact that these components are much less dominant at the recombination epoch. These constraints were obtained with only the high-$l$ ($l \geq 30 $) and CMB-\textit{lensing} \cite{Planck18_parameters} likelihoods and appropriately reinterpreting the relevant parameters. The implemented prior is a combination of physical densities of cold dark matter, baryons and the expansion rate at recombination $H^{\rm rec}$ and are summarized in \Cref{tab:CMB-E15} along with the covariances\footnote{These covariances are constructed from the covariance matrix provided by \citet{Verde17}.}. The redshift of recombination in our analysis is effectively fixed to $z_{rec} = 1089.0$\footnote{Note that we interchangeably use the notation $z_{\rm rec}$ and $z_{*}$, for ease of comparison with the earlier analysis, yet are the same quantities.}, as suggested in \cite{Verde17}, given that the relative error is of the order $\sim 0.05\%$ and would yield no major difference to the inferred constraints. Originally, the authors of Ref.~\cite{Verde17}, have suggested the implementation of priors in the form of dimensionless matter densities at recombination $\{\Omega_{\rm b}^{\rm rec},\Omega_{\rm c}^{\rm rec}, H^{\rm rec}\}$, however this also implies that the constraints on the physical densities $\{\Omega_bh^2,\Omega_ch^2\}$\footnote{The physical densities are independent of the actual $H_0$ values as for the definition of  critical density ($\rho_{c0}= 3 H^2_0 m^2_{pl}$), used to obtain the dimensionless densities ($\Omega_{i}$).} are essentially invariant, which we verify to be true across models and implement here as priors. These priors in fact are complementary to the $\Lambda$CDM model, as the necessary initial conditions at recombination and we choose this combination of data to quote our main inferences for the model. The inclusion of this early universe CMB prior in the analysis is indicated as `EUp-15'. {We replicate the analysis in \citet{Verde17} for which the covariance are reported in \Cref{sec:Appendix} and elaborated in \Cref{sec:Results}. Inclusion of the $Planck$ 2018 early-Universe priors are indicated in the analysis as `EUp-18'.}
We denote the analysis performed through this methodology using only the early-time information within the CMB data as 'Early Universe Analysis' as `EUA'.

{\renewcommand{\arraystretch}{1.5}%
    \centering

\begin{table}[!ptb]
   \caption{Mean values and the corresponding covariances of the EUp-15 (early Universe priors from {\it{Planck15}}). $H_{\rm rec} $ and $r_{\rm d}$ are reported in $\textrm{Mpc}^{-1}$ and $\textrm{Mpc}$, respectively.}
   \label{tab:CMB-E15}
    \begin{tabular}{ccccccc}      
        \hline
        \hline
        Observable & Mean & $\sigma_i$ &  \multicolumn{4}{c}{$r_{ij}$} \\ 
        \hline
$10^2 \Omega_b h^2$   & 2.227 & 1.77 $\times 10^{-2}$ & 1. & -0.68 & -0.66 & 0.33 \\
$10^2 \Omega_c h^2$   & 11.90 & 1.74 $\times 10^{-1}$ & -0.68 & 1. & 0.99 & -0.92 \\
$H^{\rm rec}$   & 5.188 & 2.44 $\times 10^{-2}$  & -0.66 & 0.99 & 1. & -0.93 \\
$r_d $   & 147.49 & 0.36 & 0.33 & -0.92 & -0.93 & 1. \\

\hline
\hline
    \end{tabular}

\end{table}
}

\textit{Reduced CMB likelihoods} (CMB/CMB [high-$l$]): We use the reduced CMB likelihoods through the distance priors simply by constructing the covariance among the observables $\{R(z_{*}), l_A(z_{*}), \Omega_bh^2\}$ \citep{Wang06, Wang07}. Here we fix the redshift $z_{*} = 1089.79 \pm 0.26$ to its mean value, given the small value of the relative error$\sim 0.02\%$. We indeed verify that the inferences are equivalent and unchanged, when the $z_{*}$ is either fixed or computed numerically, or even when utilizing the fitting formula from \cite{HuSugiyama95}. The uncertainty/covariance matrix for these observables was constructed utilizing the publicly available Planck chains\footnote{We use the \textit{Planck} $w$CDM chains obtained using complete dataset combination of high-$l$+low-$l$+lowE+$lensing$ denoted as `{\texttt{TTTEEE$\_$lowl$\_$lowE$\_$post$\_$lensing}}' in the publicly available chains.} for the $w$CDM cosmology, which we represent in the analysis simply as `CMB', unless otherwise mentioned. 

% \href{http://pla.esac.esa.int/pla/#home}

Additionally, we also implement a second reduced distance priors likelihood, in the spirit of the EUA, by first constraining the $w$CDM model utilizing only the high-$l$ ($l\geq30$) \texttt{TTTEEE} and the CMB-$lensing$ likelihoods from the \textit{Planck} 2018 release. Note that the aforementioned EUp are derived using the \textit{Planck} 2015 likelihoods. Inclusion of these distance priors using only the high-$l$ CMB likelihoods constraints are represented in the analysis as `CMB [high-$l$]'. Constraints from this CMB analysis are presented in \Cref{sec:Results} and reduced likelihoods for both these CMB dataset combinations are reported in \Cref{sec:Appendix}. A posteriori we also conclude that the reduced likelihood estimated for the $w$CDM models performs equally well for the $\Lambda$CDM model in the joint analysis. 

\begin{figure*}[!pth]
    \centering
    \includegraphics[scale=0.8]{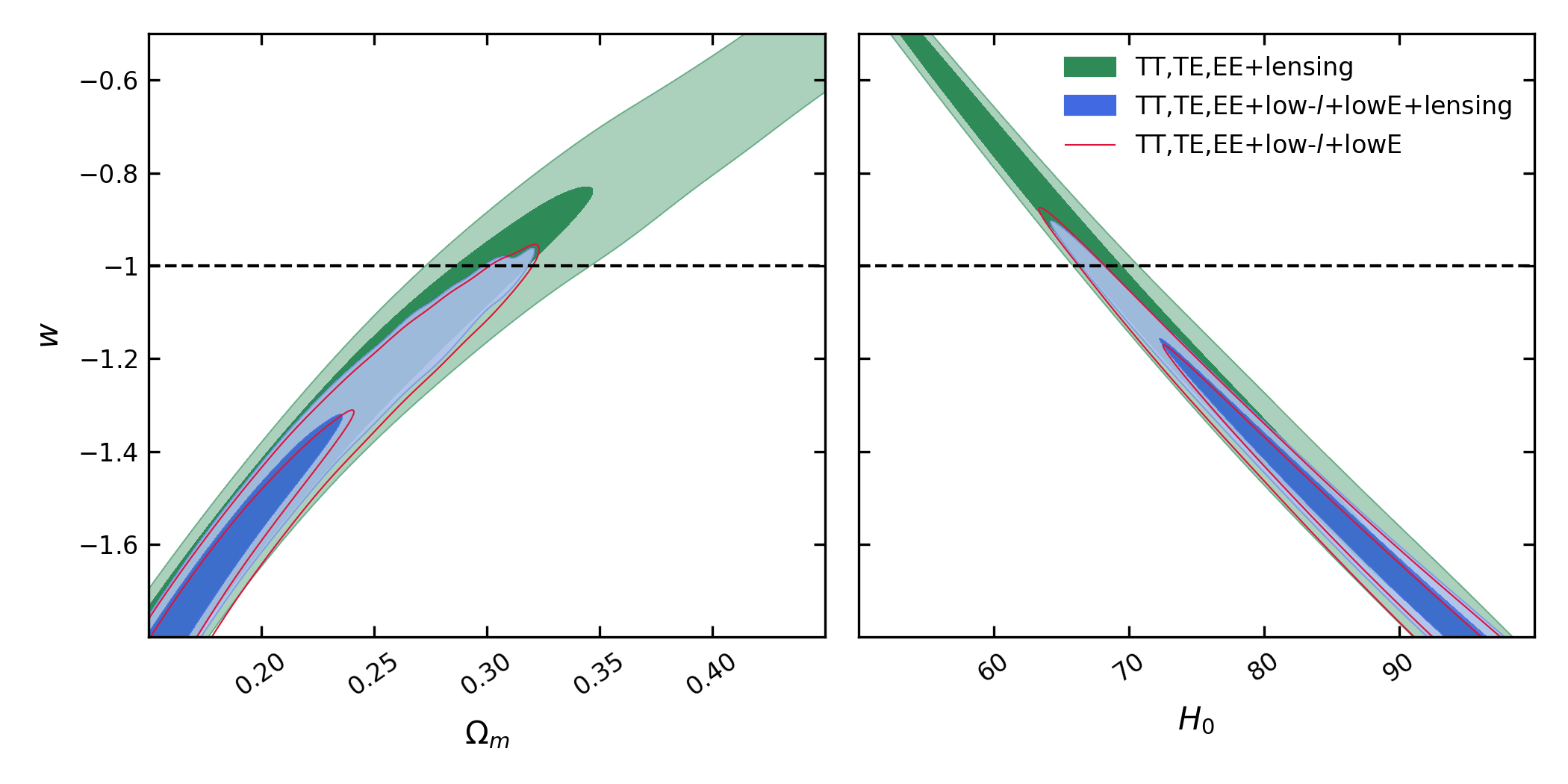}

    \caption{Comparison of the \textit{Planck} constraints for the $w$CDM model: in blue, the complete TT,TE,EE+low-$l$+lowE+$lensing$ (CMB), in green TT,TE,EE+$lensing$ (CMB [high-$l$]) combination of likelihoods and in red the combination of high-$l$+low-$l$+lowE are shown. $H_0$ is reported in the units of \ksM. The contours represent the $68\%$ and $95\%$ probability, respectively.}
    \label{fig:CMB_wCDM}
\end{figure*}

We implement the Bayesian analysis for the \textit{Planck} likelihoods using the \texttt{MontePython} package\footnote{\href{https://github.com/brinckmann/montepython_public}{https://github.com/brinckmann/montepython$\_$public}} \cite{Brinckmann18, Audren12}. We use the \texttt{emcee}\footnote{\href{http://dfm.io/emcee/current/}{http://dfm.io/emcee/current/}} \citep{Foreman-Mackey13} package, to perform the joint analysis of the low-redshift high-redshift likelihoods/priors. We then use the \texttt{getdist}\footnote{\href{https://getdist.readthedocs.io/}{https://getdist.readthedocs.io/}} package \cite{Lewis19} to analyze and infer posteriors from the chains. We also verify the comparison of the high-redshift observables between \texttt{CAMB}\footnote{\href{https://camb.readthedocs.io/en/latest/}{https://camb.readthedocs.io/en/latest/}} \cite{Lewis_2002, Lewis11} based \textit{Planck} chains and our runs which use \texttt{CLASS}\footnote{\href{https://github.com/lesgourg/class_public}{https://github.com/lesgourg/class$\_$public}} \cite{Lesgourgues11, Blas11} code. For more details on the comparisons of \texttt{CLASS} vs. \texttt{CAMB} implementations, please refer to \cite{Verde17, Lesgourgues11a}. Throughout the analysis we implement uniform, sufficiently wide flat priors on the MCMC parameters: $ 0.01 \leq \Omega_m \leq 0.5$, $50.0 \leq H_0 \leq 100.0$, $ -2.5 \leq w \leq 0.5$. When including the CMB priors we use the priors of $0.01 \leq \Omega_{\rm cdm} \leq 0.45$ and $0.02 \leq \Omega_b \leq 0.06$, on the energy densities of cold dark matter and baryons, respectively. {For the additional parameters of the EDE we implement the priors as were used in \citet{Hill20}: $0.001 \leq f_{\rm EDE} \leq 0.5$, $3.1 \leq \log_{10}(z_{\rm c})\leq 4.3$ and $0.1 \leq \theta_{i} \leq 3.1$. Unless otherwise mentioned, we assess the MCMC analysis as converged with the Gelman-Rubin criteria \cite{Gelman92} being at least $R-1 <0.01$. Note that we sample on the effective parameters for the EDE model with uniform priors. \citet{Hill20} however show that implementing flat uniform priors on the physical parameters of the scalar field in fact provide tighter limits on the effective parameters, therefore having conservative final limits.}

\section{Results and discussion}
\label{sec:Results}

We begin by contrasting the CMB constraints on the $w$CDM model, all the \textit{Planck} 2018 likelihoods and the high $l$ datasets alone, as described in \Cref{sec:Data}, essentially finding that exclusion of the low multipoles relaxes the constraints on $w$ and $H_0$. As shown in the right panel of \Cref{fig:CMB_wCDM}, the CMB constraints from full dataset, which indicate much higher values\footnote{Note that this $w$CDM based constraint on $H_0$ from \textit{Planck} agrees with the local measurement of $H_0 = 73.48\pm 1.42 \, \ksM$ reported in \cite{Riess18_H0} within $2\sigma$, owing to the $95\%$ C.L. limit of $H_0>70.4\, \ksM$.} of $H_0>82.4 \,\ksM$ at $68\%$ C.L. are relaxed to $H_0 = 78^{+10}_{-20} \, \ksM$. At the same time the phantom-like constraints of $w = -1.57^{+0.16}_{-0.33}$ from CMB are also relaxed to a $w-1.30^{+0.38}_{-0.54}$ at $68\%$ C.L. limits. One can also infer that the exclusion of the low-$l$+lowE likelihoods allow for a larger value of $H_0$ for the $w=-1$ ($\Lambda$CDM) parameter space from the same comparison in the \Cref{fig:CMB_wCDM}. In \Cref{tab:wCDM_CMB}, we report the constraints on the six base parameters and the additional EoS parameter $w$. Note the difference in reporting the parameter $100\theta_{\rm s}$ in our analysis of CMB [high-$l$] to $100\theta_{\rm MC}$ in \textit{Planck} CMB  chains, due to varied implementation of the same from the \texttt{CLASS} to \texttt{CAMB} code, respectively.

{\renewcommand{\arraystretch}{2.}%
    \setlength{\tabcolsep}{7pt}%
\begin{table}[ht!]
    \centering
    \caption{Comparison of the $68\%$ C.L. limit constraints for the $w$CDM model from the complete CMB (second column) and with the exclusion of low-$l$+lowE (CMB [high-$l$]) likelihoods. In the last two rows, we show the derived$^{*}$ $H_0$ constraints in the units of $\ksM$ and $\sigma_8$, respectively.}
    \label{tab:wCDM_CMB}
    \begin{tabular}{cccc}
        \hline
        \hline
        Parameter & CMB [high-$l$] & CMB\\
        \hline
        {$10^{-2}\omega_{\rm b }$} & $2.247\pm 0.016         $ & $2.243\pm 0.015        $\\

        {$\omega_{\rm cdm }  $}    & $0.1188\pm 0.0015          $ & $0.1193\pm 0.0012          $\\

        {$100\theta_{\rm s/MC } $} & $1.04199\pm 0.00030     $ & $1.04099\pm 0.00031        $\\

        {$\ln (10^{10}A_{\rm s })$}    & $3.085^{+0.032}_{-0.048}   $ & $3.038\pm 0.014            $\\

        {$n_{\rm s }         $}    & $0.9673\pm 0.0052          $ & $0.9666\pm 0.0041          $\\

        {$\tau_{\rm reio }   $}    & $0.076^{+0.018}_{-0.025}   $ & $0.0523\pm 0.0074          $\\

        {$w_0            $}    & $-1.30^{+0.38}_{-0.54}     $ & $-1.57^{+0.16}_{-0.33}     $\\
        \hline
        
        $H_0^{*}         $     & $78.0^{+10.0}_{-20.0}            $ & $> 82.4                    $ \\
        
        $\sigma_8^{*}           $    & $0.91\pm 0.11     $ & $ 0.964^{+0.090}_{-0.044}     $\\
        \hline
        \hline
    \end{tabular}
\end{table}
}

As expected, the constraints on the high-redshift behavior remain unchanged when excluding the low-$l$+lowE likelihoods. We find the sound horizon at the drag epoch to be $r_{\rm d} = 147.29\pm 0.32\,{\rm Mpc}$ and $z_{*} = 1089.69 \pm 0.31$, which are practically equivalent to the constraints from complete CMB dataset. This reasserts the `early universe constraints' analysis in \citetalias{Verde17}, at the same time hinting possible modifications of low-$l$ modeling, that might aid to alleviating the $H_0$-tension. The distance priors based reduced likelihood for CMB [high-$l$] constraints are presented in \Cref{sec:Appendix}. In summary, the major effect of excluding the low-$l$+lowE likelihoods is seen as degradation of the constraining ability of CMB data on the parameters of late-time physics ($H_0\,, w$), including the reionization optical ($\tau_{\rm reio}$) depth, which is however the least constrained parameter within the CMB analysis. The values of reionization optical  depth are pushed towards higher values $\tau_{\rm reio} = 0.076^{+0.018}_{-0.025}$ with larger uncertainty, yet in agreement with the tight constraint of $\tau_{\rm reio} = 0.059\pm 0.006$, reported in \citet{Pagano19}, within standard analysis. 

\begin{figure*}[!ht]
    \centering
    \includegraphics[scale=0.8]{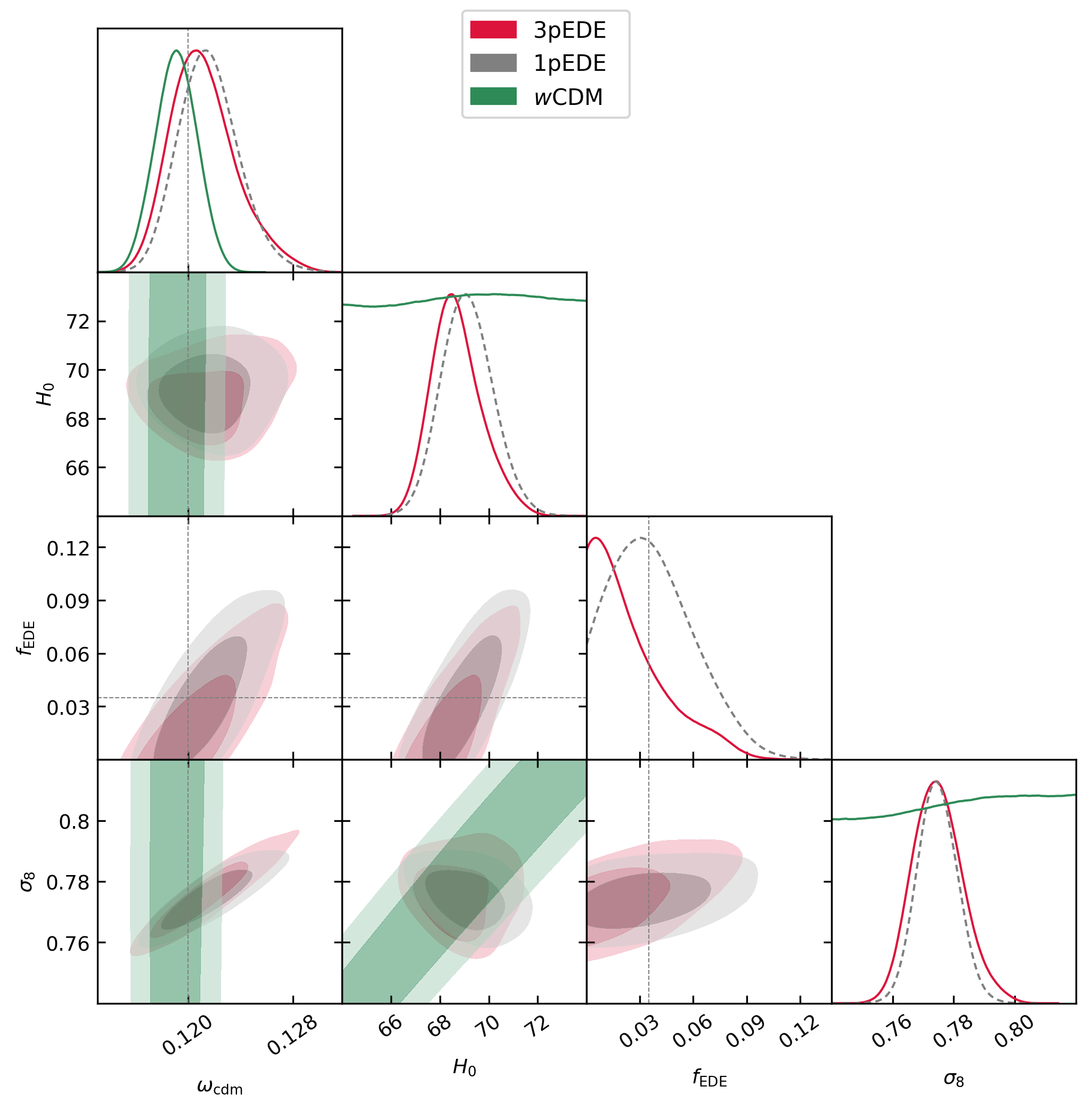}
    \caption{Early Universe Analysis (EUA) comparison of the $w$CDM model and the EDE model. The 3pEDE $95\%$ C.L. limits corresponds to $f_{\rm EDE} < 0.07$. The 3pEDE and 1pEDE analysis have the convergence criteria of $R-1\lesssim 0.08$ and $R-1\lesssim 0.01$, respectively. The dashed markers show the best-fit values of the $f_{\rm EDE}$ and $\omega_{\rm cdm}$ parameters in the 3pEDE model.}
    \label{fig:EUA}
\end{figure*}

We replicate the analysis performed in \citetalias{Verde17}, which essentially excludes the low-$l$+lowE likelihoods and marginalizes over the amplitude and tilt parameters of the Newtonian lensing potential (see Eq 4.1 in \citetalias{Verde17})\footnote{Please refer to \citet{Vonlanthen10, Audren12} for earlier discussion on the procedure to disentangle the late-time effects and the early universe physics in the CMB analysis.}, which are now additional parameters of the MCMC analysis. As implemented in \citetalias{Verde17} the reionization optical depth is fixed to $\tau = 0.01$, owing to the degeneracy with the parameter amplitude of the scalar fluctuations $A_s$. In this scenario, the extrapolated $H_0$ is not interpreted as the current expansion rate, but rather fixes the distance to the last scattering surface. Note that in \Cref{fig:wCDM_EUp} (Appendix A), we show the comparison of the constraints on the matter density and expansion rate at recombination between the \textit{Planck} 2015 analysis in \citetalias{Verde17} and our \textit{Planck} 2018 analysis, finding no major difference. The mild differences are within the correlations between the physical densities and a small shift in the constraint on $\omega_b$, as shown in \Cref{tab:CMB-E18}. 

{\renewcommand{\arraystretch}{1.75}%
    \setlength{\tabcolsep}{7pt}%
\begin{table*}[!ht]
    \centering
    \caption{Constraints for the $w$CDM and EDE models cosmological models using the CMB high-$l$ (TT+TE+EE) and lensing likelihoods in the Early Universe Analysis. The reported constraints are $95\%$ C.L. limits. Derived quantities are highlighted with a $^*$ superscript. The $\chi^2$ improvement from $w$CDM to 3pEDE is $\Delta\chi^2\sim 1$, with $\chi^2_{w \rm{CDM}} = 2355$.  }
    \label{tab:EUA}
    \begin{tabular}{cccc}
    \hline
        Parameter& $w$CDM & \multicolumn{2}{c}{EDE [n = 3]}  \\ 
        & & 3p & 1p \\ 
        \hline
        \hline
        {$10^{-2}\omega_{\rm b }$} & $2.246^{+0.035}_{-0.035}   $ & $2.260^{+0.043}_{-0.041}   $ & $2.273^{+0.046}_{-0.044}   $ \\

        {$\omega_{\rm cdm }  $} & $0.1190^{+0.0032}_{-0.0033}$ & $0.1211^{+0.0053}_{-0.0046}$ &  $0.1216^{+0.0046}_{-0.0044}$\\

        % {$100\theta_{s } $} & $1.04201^{+0.00060}_{-0.00060}$ & $1.04174^{+0.00065}_{-0.00069}$&  \\

        $e^{-2\tau} A_{\rm s} $ & $1.878^{+0.025}_{-0.024}   $ & $1.882^{+0.030}_{-0.029}   $ & $1.918^{+0.027}_{-0.024}   $ \\

        {$n_{\rm s }         $} & $0.966^{+0.011}_{-0.010}  $ & $0.971^{+0.015}_{-0.014}   $ & $0.976^{+0.015}_{-0.013}   $ \\

        {$w^0_{\rm fld }      $} & $-1.27^{+0.72}_{-0.69}   $ & -- & -- \\
        {$\log_{10}(z_{\rm c })    $} & -- & $3.72^{+0.56}_{-0.45}    $ & -- \\

        {$f_{\rm EDE}    $} & -- & $< 0.071            $ & $< 0.080                  $ \\

        {$\theta_{\rm i}  $} & -- & $2.0^{+1.1}_{-1.7}       $ & -- \\
        \hline
        {$z_{\rm rec }^{*}      $} & $1088.96^{+0.65}_{-0.71}   $ & $1089.02^{+0.67}_{-0.65}   $ & $1088.89^{+0.66}_{-0.60}   $\\ 
        {$r_{\rm d }^{*}     $} & $147.24^{+0.63}_{-0.63}    $ &  $145.9^{+1.7}_{-2.6}       $ & $145.5^{+2.0}_{-2.3}       $ \\ 
        % \hline
        % $\chi^2_{\rm b.f}$ & 2355 & 2354 \\ 

        \hline
        \hline
    \end{tabular}
    \end{table*}
}
 
In this context we also perform the Early Universe Analysis of \citetalias{Verde17}, as described above, with the EDE model. Using the same CMB data along with the additional three EDE parameters ($\Theta_{\rm EDE}$) we assess the evidence for the EDE model. We find a limit of $f_{\rm EDE} < 0.07$ at $95\%$ C.L., which is similar to the limits of $f_{\rm EDE} < 0.06$ \cite{Hill20} found with the complete CMB data and additional low-redshift large scale structure data, without the inclusion of local $H_0$. In fact, our $95\%$ C.L. from the EUA is slightly tighter than $f_{\rm EDE} < 0.087$ obtained from complete CMB data alone. Our limit from EUA is also equivalent to the limits found in Ref.~ \cite{Hill20} with the complete CMB data, BAO, SN and Full Shape analysis of the BOSS power spectrum. This essentially indicates that the there is an even weaker evidence for the EDE modification, when assessing it only against the early-time information in the CMB data. Also, the $\chi^2$ improvement is of the order of $\Delta \chi^2 \sim -1.0\,  (\chi^2_{\rm EDE} = 2354)$ for the EDE modification. In \citet{Hill20}, it was reported that the $\chi^2$ for the high-$l$ (TT+TE+EE) dataset becomes worse with the EDE modification by $\Delta \chi^2 \sim 2.6$, when including the local $H_0$ value (please see Table VII of \cite{Hill20}), in the joint analysis. Also indicating that the early-time physics within the CMB data does not support the current early dark energy extension to $\Lambda$CDM. However, note that more recently in Ref.~\cite{Smith20}, it has been shown that fixing the $\{\log_{10}(z_{\rm c}), \theta_{\rm i}\}$ parameters considerably increases the allowed range for the $f_{\rm EDE}$ parameter (1pEDE), while having the same $\chi^2$ value. This implementation appropriately implies a fine-tuned early universe allowing to sample better the $f_{\rm EDE}$ parameter space, which in a MCMC analysis tend to collapse the posterior distribution on the $\Lambda$CDM model ($f_{\rm EDE} \sim 0$). Similar inference was also made in Ref.~ \citep{Lin20} for acoustic dark energy and in Ref.~\citep{Braglia20} for early modified gravity models. This highlights the dependence of the posteriors on the choice of priors, elaborated in Ref.~ \citep{Murgia20}. Within the EUA we find the best-fit values of $\{\log_{10}(z_{\rm c}), \theta_{\rm i}, f_{\rm EDE}\} = \{ 3.89, 2.74, 0.035\}$, which for the former two parameters are similar to values from the full CMB analysis \citep{Hill20, Smith20, Murgia20}. The $f_{\rm  EDE}$ however has a lower best-fit value, in comparison to the 0.068 in Ref.~\citep{Hill20} and 0.085 in Ref.~\citep{Murgia20}. We infer these differences as the effect of the late-time information within CMB data also for the EDE model. Although the limits obtained in our analysis are tighter, in agreement with Ref.~\citep{Murgia20}, we also find that our best-fit  $f_{\rm EDE}$ value is different from the peak of the posterior and close to the $1\sigma$ bounds. 

We stress that one should interpret the strengthening of the  limits, obtained with the exclusion of late-time physics in the EUA analysis, as a shift in the confidence regions, however sharply bounded on the lower end, which gives the impression of tighter limits. Tables VII-X in \citep{Murgia20} show that in the 1pEDE formalism and/or the inclusion of the local $H_0$ (\citetalias{Riess18_H0}), yielding higher values of $f_{\rm EDE}$, always worsens the fit to the high-$l$ likelihood. This validates our tighter limits (lower values) on $f_{\rm EDE}$ parameter which is obtained using only the high-$l$ information, while having a mild preference for the EDE over the $\Lambda$CDM model. From the $\chi^2$ comparison, noticing that higher values of $f_{\rm EDE}$ imply better fit to low-$l$ and worse fit to high-$l$ likelihoods we expect for the 1pEDE with the EUA lacking the low-$l$ pull, might not be able to relax the bounds on $f_{\rm EDE}$. {To Validate the same, we perform the EUA with the 1pEDE model, wherein we fix the $\{\log_{10}(z_{\rm c}), \theta_{\rm i}\}$ parameters to their best fit from the 3pEDE analysis. We find that the mean of the posterior now shifts to be in better agreement with the best-fit of $f_{\rm EDE}$ from the the 3pEDE analysis. The $95\%$ upper limit also increases mildly to $< 0.08$ with a $65\%$ C.L. constraint of $f_{\rm EDE} = 0.038^{+0.015}_{-0.031}$.}

In addition to the points made above, Ref.~\citep{Smith20} shows that lower values of $A_{\rm s}$ obtained from the LSS observations, in comparison to the CMB constrains lead to the discrepancy claimed by \citep{Ivanov20}. Within our EUA, we do not have constrains on the $A_{\rm s}$ parameter alone but only for the combination parameter $e^{-2\tau}A_{\rm s}$. We find that the quantity $e^{-2\tau}A_{\rm s}$ is almost equivalent for both the models and is also compatible with the $\Lambda$CDM constraints obtained from the full data \cite{Planck18_parameters}. In \Cref{fig:EUA}, we show the comparison of the constraints on $H_0$ and $\sigma_8$ in the EUA for the two models. We stress that in the EUA there exists no interpretation of $H_0$ and $\sigma_8$, which are parameters constraining late-time physics. However, the strong positive correlation between the $H_0$ and $\sigma_8$ shows why a simple dark energy extension of $w\neq -1$ is inadequate to resolve the $H_0$-tension. Higher values of $H_0$, that imply higher values of $\sigma_8$, will be in significant tension with the lower valued $\sigma_8$ constraints from the LSS \cite{Abbott17} data.

{\renewcommand{\arraystretch}{2.}%
    \setlength{\tabcolsep}{7pt}%
\begin{table*}[!ht]
    \centering
    \caption{Constraints on the $w$CDM model at 68\% confidence level obtained for various combinations of datasets. We quote the maximum posterior and the $16^{\rm th} , 84^{\rm th}$ percentiles as the uncertainty\footnote{We quote the maximum posterior values instead of the mean, as it captures the nature of skewed distributions better.}. }
    \label{tab:wCDM_params}
    \begin{tabular}{cccc}
        \hline
        \hline
        Parameter & $\Omega_m$ & $w$ & $H_0 \, [\ksM]$ \\ 
        Data & & & \\

        \hline
		SL & $0.30^{+0.13}_{-0.10}$ & $-2.32^{+0.77}_{-0.17}$ & $81.4^{+5.3}_{-5.0}$ \\ 
		SL+CMB & $0.236^{+0.022}_{-0.016}$ & $-1.315^{+0.111}_{-0.091}$ & $77.6^{+2.7}_{-3.4}$ \\ 
		SL+SN & $0.361^{+0.056}_{-0.068}$ & $-1.13^{+0.19}_{-0.24}$ & $75.1^{+2.2}_{-2.5}$ \\ 
		SL+SN+EUp-15 & $0.266^{+0.014}_{-0.010}$ & $-0.929^{+0.045}_{-0.057}$ & $72.5^{+1.7}_{-1.5}$ \\ 
		 
		SL+SN+EUp-18 & $0.268^{+0.012}_{-0.013}$ & $-0.939^{+0.049}_{-0.050}$ & $72.5^{+1.7}_{-1.6}$ \\
		SL+SN+CMB & $ 0.296^{+0.009}_{-0.010}$ & $-1.046^{+0.032}_{-0.040}$ & $69.5^{+1.1}_{-1.0}$ \\ 
		SL+SN+CMB [high-$l$] & $ 0.288^{+0.012}_{-0.009}$ & $-1.032^{+0.039}_{-0.035}$ & $69.6^{+1.2}_{-1.0}$ \\
				\hline
        \hline
    \end{tabular}
\end{table*}
}

We now proceed to perform the joint analysis of the low-redshift SN and SL datasets in conjunction with the CMB/EUp priors as described in \Cref{sec:Data} and reported earlier in this section. In \Cref{tab:wCDM_params}, we show the constraints for the $w$CDM model parameters. As it can be seen the first two rows are in very good agreement with those quoted in \citet{Wong19}, except for mild differences in the posteriors. We also implement slightly different uniform priors on the parameters as reported in \Cref{sec:Data}, with no consequence for the inferred posteriors. Considering the dataset SL+SN+CMB for the $w$CDM model, the posterior on $H_0$ is driven towards lower values, due to the almost orthogonal correlations in the $\Omega_{\rm m}\, \textrm{vs.}\, w$ plane, of the CMB and SN+SL datasets. Replacing the CMB distance priors with CMB [high-$l$] priors induces a parallel shift (keeping the correlation intact) in the contours towards $w>-1$ and mildly higher values of $H_0$. The variation in the CMB constraints shown in \Cref{fig:CMB_wCDM} with the exclusion of low-$l$+lowE datasets, does not weigh enough to modify the $H_0$ estimates within the $w$CDM framework. The $H_0$-tension here is reduced to $\sim 2.2\sigma$ for both the CMB and CMB [high-$l$] priors, with mildly phantom-like dark energy EoS. While we did not show the constraints on $r_d$ here, we verify that they are in agreement with the expectation reported in \Cref{tab:CMB-E15,tab:CMB-E18}, when early universe priors are included. 

Through the parallel shift of the contours from CMB to CMB [high-$l$] priors, we notice that along the $w=-1$ parameter space in the \Cref{fig:wCDMe}, the constraints on the $H_0$ value are shifted towards higher values and with mildly larger uncertainty, which aids the supposition that with the exclusion of low-$l$ likelihoods the tension can be moderately alleviated, however not adequate. The constraint on Hubble constant in the $\Lambda$CDM scenario shifts from $H_0 = 68.24^{+0.55}_{-0.61}\,\ksM$ to $H_0 = 68.97^{+0.77}_{-0.64} \,\ksM$, corresponding to $\sim 3.4\sigma$ and $\sim 2.8\sigma$ tension, respectively. 

\begin{figure}[!ht]
    \centering
    \includegraphics[scale=0.5]{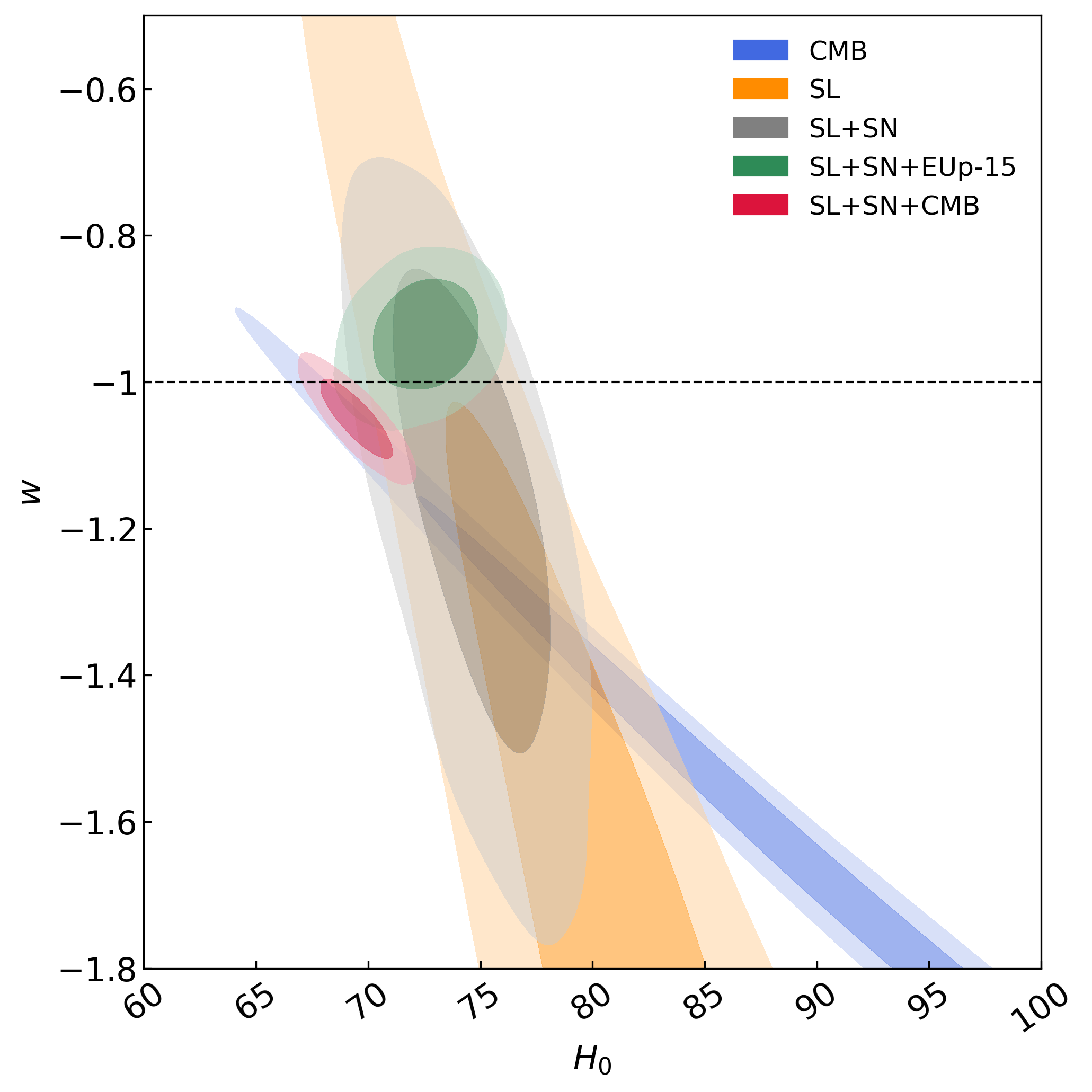}
    \caption{Constraints from various dataset combinations within the $w$CDM model ($H_0$ has units of \ksM). Here the CMB data implies complete analysis of the TT+TE+EE+low-$l$+lowE+lensing datasets for the $w$CDM model.}
    \label{fig:wCDM}
\end{figure}

At the same time, analysis with the inclusion of the EUp-15 allows for a larger range of possible $H_0$ value in agreement with the local \citetalias{Riess18_H0} estimate and notably for quintessence-like EoS $w>-1$, owing to the modified correlations in $H_0\,\textrm{vs.}\, w$ parameter space (see also \Cref{fig:wCDM}). Clearly, the observables through which the priors are implemented play a major role in determining the final correlations in the joint analysis. And this might even be a more appropriate analysis in comparison to the inclusion of the prior using the shift parameters, to asses the effects of excluding low-$l$+lowE likelihoods and combining the early CMB physics and late SN dynamics. Indeed, the distance priors as in \Cref{eq:Distance_Priors} also take into account the late-time behaviour of the given model through the angular diameter distance $D_{\rm A}(z^{*})$. As shown in \Cref{fig:wCDMe}, simple exclusion of the low-$l$+lowE likelihoods shifts the constraints in the direction of the EUp-15/18, and further excluding the effects of late-time physics on the high-$l$ part of the CMB spectrum, moves the EUp-15/18 constraints completely alleviating the $H_0$-tension with the local \citetalias{Riess18_H0} estimate. Both the `EUp-15' and `EUp-18' priors provide identical constraints and are in equally good agreement\footnote{Note that the constraint obtained from Hubble rate dataset, namely Cosmic Chronometers \cite{Jimenez02, Moresco12,Moresco15,Moresco16}, which is $H_0 = 68.52^{+0.94+2.51({\rm sys})}_{-0.94}\,\ksM$ \cite{Haridasu18_GP}, will be in agreement with this estimate when accounting for the systematics.} with the \citetalias{Riess18_H0}, being $H_0 = 72.5^{+1.7}_{-1.6}\, \ksM$. As shown in \Cref{fig:wCDM_EUp} (see also \Cref{tab:CMB-E18}), the mild shift in the priors does not provide any visibly distinguishable effects in the joint analysis. This is clearly in accordance with the fact that the major improvements from the \textit{Planck} 2015 to \textit{Planck} 2018 likelihoods is within the modeling of the low-$l$ and lowE likelihoods \cite{Planck18_parameters, Planck16}, essentially keeping the early universe constraints unchanged. As shown in \citetalias{Verde17}, the early universe priors would remain unchanged when a $\Omega_{\rm k} \neq 0$ ($k\Lambda$CDM) freedom is taken into account, instead of the $w\neq -1$, freedom. This in turn asserts that our inferences here would be unaffected, even if the entire analysis is performed again using the $k\Lambda$CDM model.

\begin{figure}[!ht]
    \centering
    \includegraphics[scale=0.50]{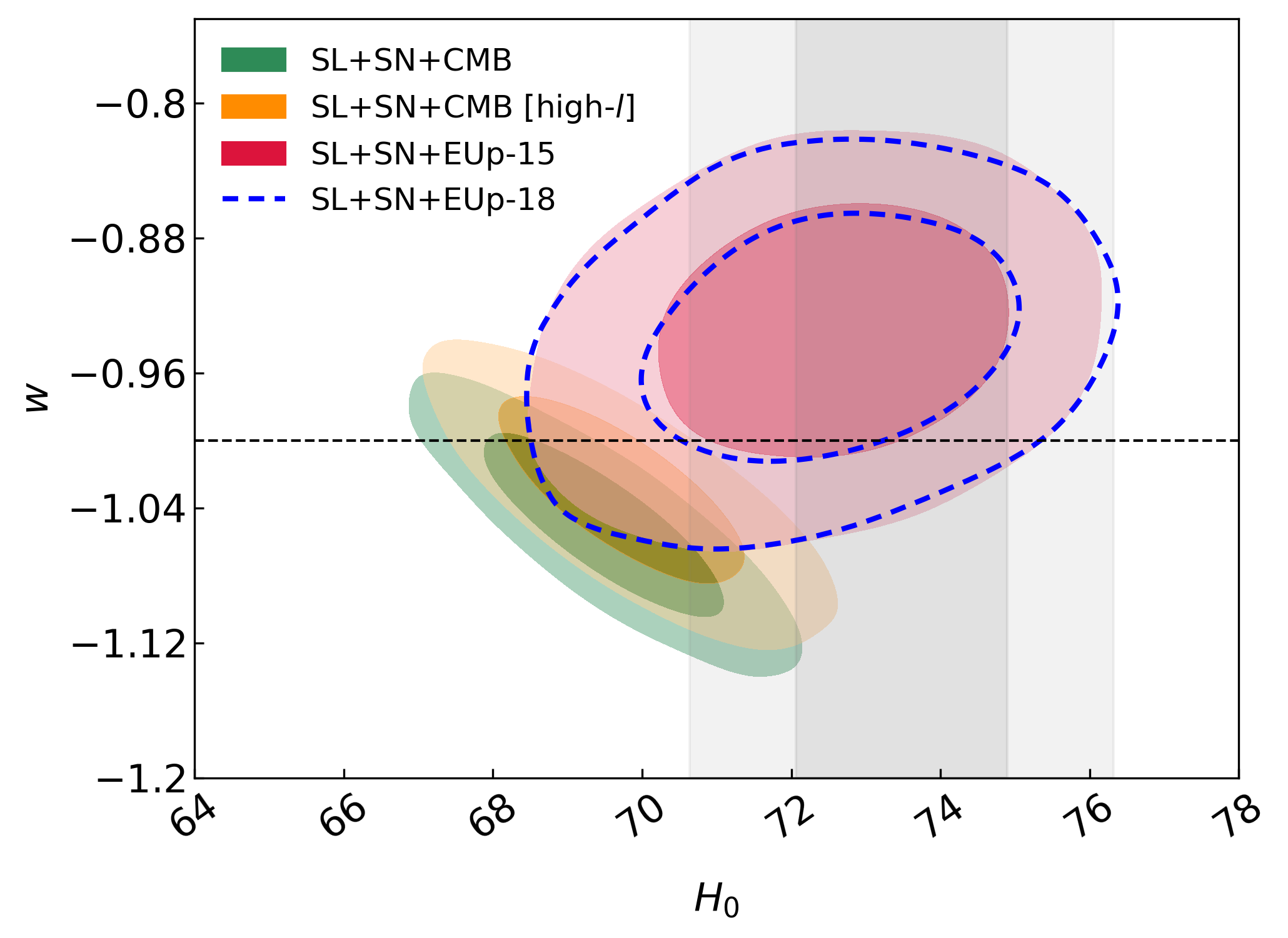}
    \caption{We show the comparison between the the final $H_0$ constraints for the three CMB priors used in the analyses. The gray band represents the local $H_0$ estimate of \citetalias{Riess18_H0}.} 
    \label{fig:wCDMe}
\end{figure}

The $H_0 = 72.2\pm 1.6\, \ksM$ constraint for the $\Lambda$CDM ($w=-1$) is also very much in agreement with the local \citetalias{Riess18_H0} value. This tentatively indicates that the early universe physics needs not be modified while seeking a resolution to the $H_0$-tension, and in turn the low-$l$ likelihoods and the late-time effects on the high-$l$ modeling of the CMB datasets could be appropriately revised. The $H_0$-tension which is usually assessed as a tension between the early physics constrained by CMB and late-time local estimate, we now recast as a disagreement between the late-time physics affecting the CMB analysis and the local estimate. As it is elaborated in Refs.~\citep{Vonlanthen10, Audren12, Audren14, Verde17}, the late-time cosmology effects enter the CMB analysis through: $i)$ late integrated Sachs Wolfe effect (ISW), which affects the low-$l$ ($l<30$); $ii)$ suppression of amplitude for $l \gg 40$, through the reionization effects and $iii)$ estimate of angular diameter distance ($D_{\rm A}(z_{\rm rec})$) to recombination, which in turn affects the angular scales\footnote{The angular scale, i.e., location of the acoustic peaks in the CMB spectrum,  essentially constraints the combination of the sound horizon ($r_s(z_{\rm rec})$) and the angular diameter distance at the recombination, which are early-time and late-time physics dependent, respectively. The assumption of the same as fiducial cosmology in obtaining the BAO observables in turn makes them moderately dependent on late-time CMB physics, and might prevent the BAO data from being `truly' model-independent.}. Clearly implying that these three effects of the late-time physics on high-$l$ CMB spectrum mark the difference between the constraints obtained utilizing the CMB [high-$l$] priors to the EUp-18/EUp-15 priors in our analysis. Finally, we also add the latest BAO dataset only to assert the variation in inferred value of $H_0 = 68.98^{+0.57}_{-0.80} \, \ksM$ for the $\Lambda$CDM model, with the EUp-15 priors. This estimate is equivalent to using the CMB [high-$l$] priors with out the BAO data, and might reflect the effect of assuming the low-redshift behavior thorough the fiducial cosmology, especially through the anisotropic component of BAO observables as elaborated in \citet{Ivanov19}. Excluding the BAO data from our main analysis and the final inferences is partly also due to the speculation of underestimated error bars in the traditional BAO analysis \cite{Anselmi18, Anselmi17, Anselmi17b} and possible discordance with BAO data \cite{Haridasu17_bao}, which are mild and however require further validation. 

Indeed, it is the physics of low-$l$ ($l \leq$ 30) dataset within the CMB analysis, which is not yet very well understood, owing to the low-$l$ anomalies \citep{Rassat14, Schwarz16}, physics of reionization (polarization) \citep{Billi19, Obied17}, and the constraints from the ISW effect, for example in Refs.~\citep{Das13, Velten15, Mostaghel18, Soltan19}, it is claimed that some level of deviation form the $\Lambda$CDM model, in different physical scenarios. While we have not performed a low-$l$ only analysis, the constraints, even in the current form, are instructive that low-$l$ dataset alone would be inadequate to constrain cosmological models or at least in major disagreement with the other low-redshift probes like supernovae. Analyses and further progress in the direction of understanding the low-$l$ physics could be thus promising.

It is also informative to contrast the early universe constraints to the simple jackknife-like split of CMB data at $l \sim 800$, as shown in \cite{Planck18_parameters, DiValentino19}. In the $\Lambda$CDM scenario, this split indicates that the a lower value of $H_0$ (higher value of $\Omega_{\rm c}h^2$) is preferred for $l>800$ (with \texttt{TT,TE,EE} dataset), indicating a mild discrepancy. This is removed when the $lensing$ data is included (see Fig. 22 of \cite{Planck18_parameters}) and the constraints become consistent with $l<800$ data, emphasizing the effect/impact of lensing data at higher multipoles. Also, when marginalizing on the lensing potential, \citetalias{Verde17} report that excluding multipoles $l<200$ provide highly degraded constraints, and will indicate no discrepancy. Indeed, the variation in constraints on $\Omega_{\rm c} h^2$ (lower values from $l<800$ and higher $l>800$) would translate into a discrepancy between the $H^{\rm rec}$ estimates. This in turn would affect two distinct features: i) Silk damping at higher multipoles in the temperature and ii) the polarization spectrum for the $l\lesssim 800$ explaining the discrepancy in the constraints on parameters such as $\{H_0, \, \Omega_{\rm c}h^2\}$, when lensing data is not taken into account, in a $l\sim 800$ data split. Once the lensing dataset is taken into account (with or without marginalizing on the potential), reduces the damping making the constraints obtained using $l>800+lensing$ consistent with those from  $30<l<800$, and indicate that the discrepancy when not including lensing data might be misleading. See also \citep{Chudaykin20}, for various other data combinations, where the high-$l$ ($l>1000$) CMB \texttt{TT} and the \texttt{TE, EE} datasets are excluded, showing mild movement in the $H_0$ constraints for the $\Lambda$CDM model.

\begin{figure}[!ht]
    \centering
    \includegraphics[scale=0.54]{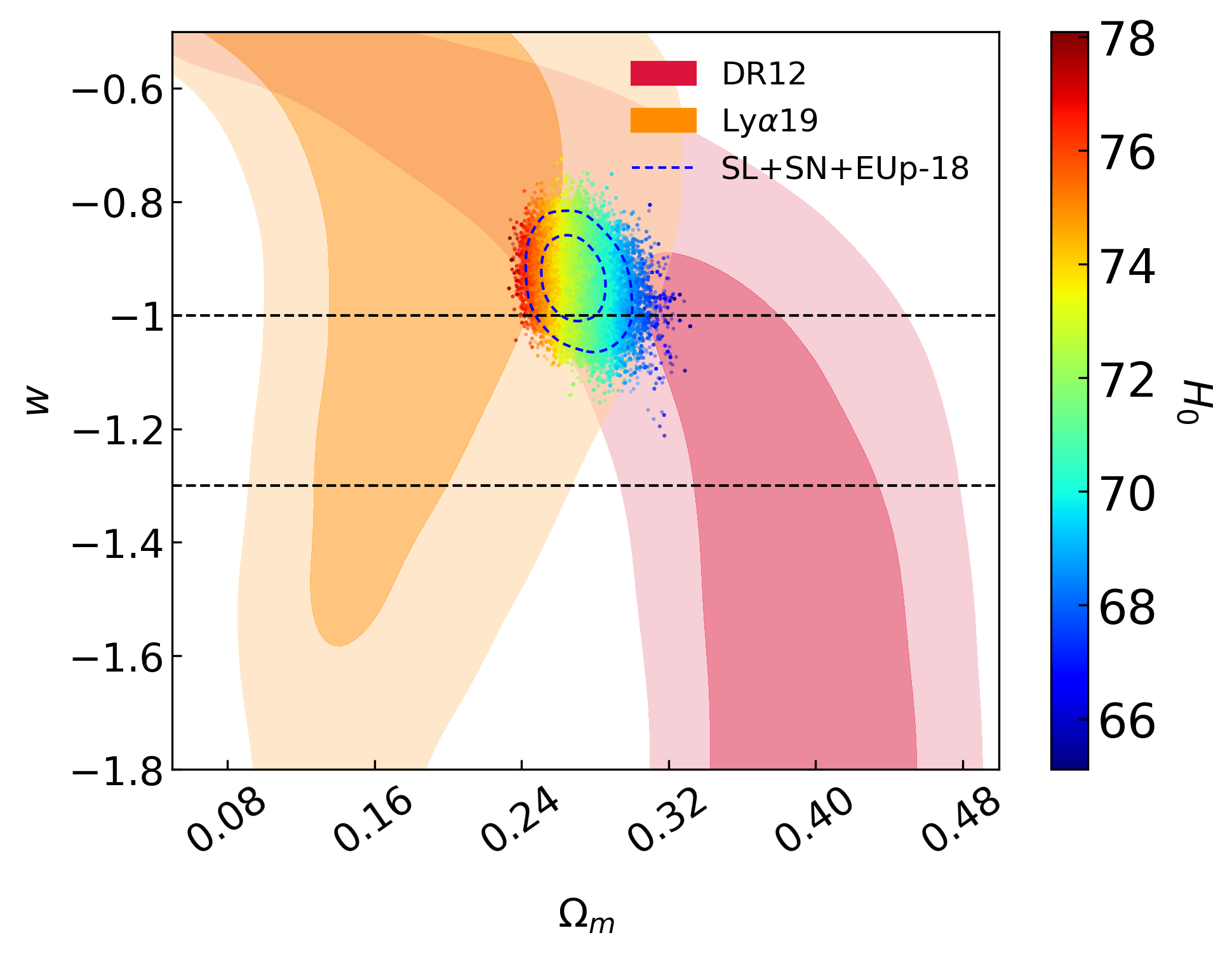}
    \caption{Comparison of the $\Omega_{\rm m}$ constraints obtained from the high-redshift Ly-$\alpha$ and the low-redshift galaxy clustering BAO DR12 datasets, for the $w$CDM model.} 
    \label{fig:BAO_wCDM}
\end{figure}

\begin{figure*}[!ht]
    \centering
    \includegraphics[scale=0.52]{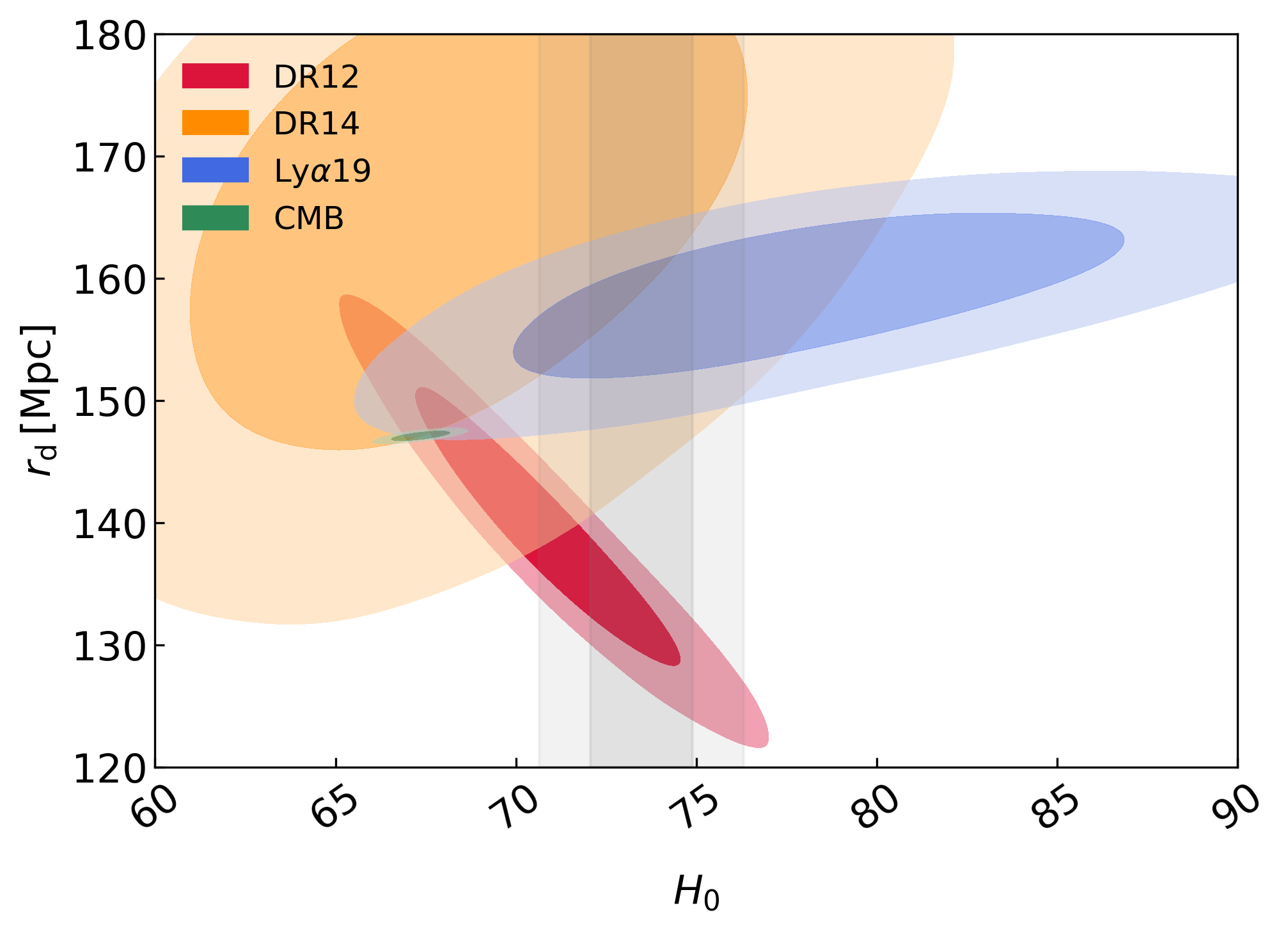}
    \includegraphics[scale=0.52]{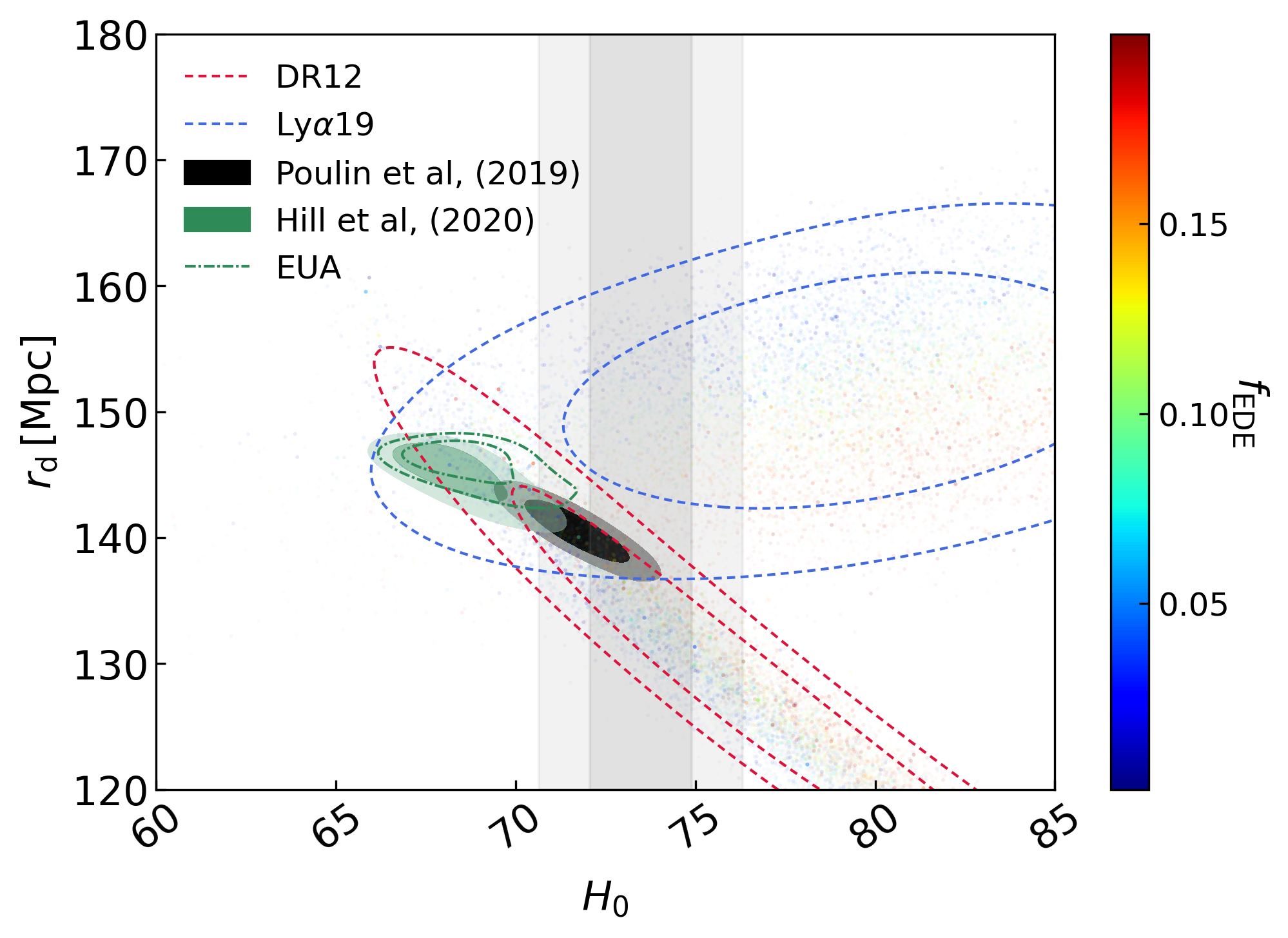}
    \caption{\textit{Left}: We show the comparison of inverse distance ladder analysis performed on different BAO datasets assuming the $\Omega_{\rm b}h^2$ prior from CMB, as shown in \Cref{tab:CMB-E15} using the $\Lambda$CDM model. Here the CMB data implies complete analysis of the TT+TE+EE+low-$l$+lowE+lensing datasets for the $\Lambda$CDM model. \textit{Right}: We show the EDE analysis with individual datasets, similar to the \textit{left} panel. The EDE contours in \textit{black} and \textit{green} are the early dark energy constraints, corresponding to $n=3$ case presented in Table II of \citet{Poulin18} and Table II of \citet{Hill20} (CMB alone), respectively. The green dashed contours show our 3pEDE constraints. Note that for brevity here we utilize the prior of $0.001\leq f_{\rm EDE}\leq 0.2$ for the BAO analysis, in contrast to the results quoted in the main text. } 
    \label{fig:BAO_LCDM}
\end{figure*}

{In \Cref{fig:BAO_wCDM}, we show the comparison of the constraints in the $w$CDM model obtained for the Ly$\alpha$19 and the DR12 datasets. The constraints on the matter density from the high-redshift Ly$\alpha$19 and low-redshift DR12 datasets are clearly less compatible for $w<-1$, which limits possibilities of any phantom-like dark energy scenarios. Also, as the scatter of the $H_0$ values within the SL+SN+EUp-18 shows, the high-redshift Ly$\alpha$19 is yet compatible with a $w\sim -1$ and larger values of $H_0$, which is not the case with the DR12 data. Including the Ly$\alpha$19 and DR12 datasets separately to the SL+SN+EUp-18 fits, we obtain $H_0 = 72.03^{+1.55}_{-1.59}\, \ksM$ and $H_0 = 69.04 \pm 0.83\, \ksM$, respectively. In this respect, one could disentangle the data corresponding to the late-time physics in CMB data and the low-redshift BAO clustering DR12 datasets as the cosmological data which is incompatible with the local $H_0$ \citetalias{Riess18_H0} estimate. Recently, in \cite{DAmico20b}, the constraints in the $w$CDM scenario have been explored with various combination of datasets. }

{In the \textit{left} panel of \Cref{fig:BAO_LCDM} we show the comparison of the inverse distance ladder analysis performed on the different BAO datasets for the $\Lambda$CDM model. We notice orthogonal nature of the correlations of the $r_{\rm d} \,vs.\,H_0$ parameters derived from the Ly$\alpha$19 and DR12 combinations, which is a known behavior and has been presented earlier (see for example \cite{Aubourg14}). However, this orthogonality implies that the usual early universe modifications, intended to reduce the sound horizon at drag epoch and hence accommodate a larger values of $H_0$ will clearly be in agreement with the DR12 dataset\footnote{The early dark energy modifications in \citet{Poulin18} modify the pre-recombination calibration of $r_{\rm d}$ as a function of the matter densities allowing for larger parameter space in $r_{\rm d}\, vs.\, H_0$ parameter space.}. And the already existing $2.3 \sigma$ tension\footnote{This tension has now been reduced to $1.5\sigma$ in the latest DR16 release \cite{Bourboux20}. However the correlation between $r_{\rm d}\, vs. \, H_0$ parameters is bound to be the same and hence the argument for the alleviated $H_0$-tension.} \cite{SainteAgathe19, Blomqvist19}, could possibly become worse or at best remain at a similar significance, and cannot {necessarily be seen as a simple statistical fluke or systematic effects in the Ly$\alpha$19 dataset}. This orthogonality of constraints in the $r_{\rm d} \, vs.\, H_0$ between the DR12 and Ly$\alpha$19 dataset might in fact suggests that an early universe modification alone cannot alleviate the $H_0$ tension and some late-time modification is also required. Also note the redshift dependent behavior of the change in the $r_{\rm d}\, vs.\, H_0$ from the DR12 ($0.2<z<0.75$), to the DR14 ($0.8<z<2.2$) and the Ly$\alpha$19 ($1.77 < z  < 3.5$) datasets. This redshift dependent behavior could be of utmost importance also with the soon to be available DR16 \cite{Hou20, Bautista20, Bourboux20} datasets, which we intend to explore elsewhere \citep{Bidenko20}.}

We also compare the constraints in the $r_{\rm d}\, vs.\, H_0$ plane for the early dark energy (EDE) analysis presented in \citep{Poulin18}, which was shown to alleviate the $H_0$-tension. However, more recently \citep{Hill20} have argued that there is no evidence for EDE  when using CMB data alone, owing to minimal improvement in the $\Delta\chi^2 \sim -4.1$ with an inclusion of 3 additional parameters, and the improvement in the $\chi^2$ is mostly contributed by the inclusion of local $H_0$ value. In \citep{Ivanov20} it has been shown that the EDE models will increase the discrepancy with the Large Scale Structure data, owing to the tension in the $\sigma_8$ parameter. This in fact, aids the motivation for the arguments raised in the current work. In Ref.~\citep{Smith20}, it has been shown that accounting for the large unconstrained distributions of $\{\log_{10}(z_{\rm c}), \theta_{\rm i}\}$ parameters can alleviate the above mentioned issues, having only one additional parameter. 

{To estimate the effect of the EDE extension on the BAO constraints, we perform a similar analysis with different BAO datasets for the EDE model. We fix the EDE parameters $\{\log_{10}(z_{\rm c}), \theta_{\rm i}\}$ as in the analysis presented in \citet{Smith20}, to the best-fit values reported in Table-II of \citet{Hill20}, varying only $\{\omega_{\rm cdm}, H_0, f_{\rm EDE}\}$. As expected, we find the contours in the $r_{\rm d}\, \rm{vs.}\, H_0$ plane, extend to the higher values of $H_0$ and lower values of $r_{\rm d}$ for both the DR12 and the Ly$\alpha$19 datasets.} However, marginalizing on the $f_{\rm EDE} \in \{0.001, 0.5\}$,  we find that the mild to moderate disagreement between the DR12 and Ly$\alpha$19 dataset remains as $\Omega_{\rm m} = 0.384^{+0.055}_{-0.062}$ and $\Omega_{\rm m} = 0.198^{+0.034}_{-0.055}$, respectively. This disagreement is at a tentative tension of $\sim 2.7\sigma$ significance. When including the local \citetalias{Riess18_H0} prior to the joint analysis of DR14 and Ly$\alpha$19 we find the constraint  $f_{\rm EDE} =0.189^{+0.065}_{-0.054}$. We present the complete contours for the parameters of the analysis of BAO datasets in \Cref{sec:Appendix2}. Conveniently, we also verify that these results agree with the $\Lambda$CDM analysis for $f_{\rm EDE} = 0$ and accordingly seen for $w = -1$ in \Cref{fig:BAO_wCDM}.

\section{Conclusions}
\label{sec:Conclusions}

In the current work, we attempt to assess the origins (necessary and sufficient modifications of physical models) of the $H_0$-tension within the CMB likelihoods, essentially distinguishing the early-time and late-time physics. We implement this through a very simple joint analysis, utilizing the CMB data in various reduced forms as priors. A summary of our primary inferences is as follows:

\begin{itemize}
    \item We find that the exclusion of the low-$l$+lowE likelihoods from the CMB analysis relaxes the phantom-like constraints on the dark energy EoS and is thus able to raise the value of high-redshift $H_0$ towards larger values being consistent with $w \rightarrow -1$. However, in conjunction with the well-constrained dynamics at low-redshift by the SNe dataset, we find it highly unlikely to be sufficient to resolve the $H_0$-tension. 
    
    \item Using early universe priors from CMB data (EUp-15/18) which disentangle the late-time and early-time physics (see also \cite{Audren12, Audren14, Verde17}), we find that the early-time physics can indeed be consistent with local estimate of $H_0$, within the $\Lambda$CDM model when excluding the galaxy-clustering BAO DR12 dataset. 
    
    \item {We analyze the EDE model using the Early Universe Analysis (EUA) of the CMB data finding no evidence for the EDE extension with $f_{\rm EDE} < 0.07$ at $95\%$ C.L. limits, sufficient to alleviate the $H_0$-tension. In the 1pEDE analysis this constraint is relaxed to $f_{\rm EDE} < 0.08$. }
    
    \item {Through our simple EUA of EDE model and due to the fact that the high-$l$ fit worsens when alleviating the $H_0$ tension in EDE model, we speculate that modifications in the physics or systematics effects might be required in the CMB low-$l$ modeling and/or late-time physics. Clearly, the analysis performed here is not sufficient to point to the actual needed physics or to the systematics, but overall implies that the late-times physics within the CMB data might be responsible for the tension.}
    
    \item {The orthogonality of the DR12 and Ly$\alpha$19 dataset constraints in the $r_{\rm d} \, vs.\, H_0$ parameter space suggests that an early universe modification alone cannot be sufficient to resolve the $H_0$-tension. At least a combination of both late and early universe modifications would be required.}
    
\end{itemize}

The inferences here, if taken at face value, recast the usual perspective of $H_0$-tension as an early-time CMB vs. local (late-time) physics effect to a late-time physics in CMB vs. local perspective. In the light of recent high precision data, the increasing $H_0$-tension is a pressing issue for the current cosmological scenario, making it very important  to be resolved as soon as possible in the near future. While newer and more precise data will aid the cause, it is vital to pinpoint the origins of the tension and find appropriate directions to drive the numerous efforts taken in the theoretical modeling and in the estimate of systematic and statistical errors.

\section*{Acknowledgments}
B.S.H and N.V acknowledge financial support by ASI Grant No. 2016-24-H.0. M.V is supported by INFN INDARK PD51 grant and agreement ASI-INAF n.2017-14-H.0. B.S.H acknowledges INFN Roma, Tor Vergata Computing Center services (RMLab) and is thankful to Federico Zani for providing assistance. We acknowledge the use of CINECA high performance computing resources under the projects `INF19\_indark\_0', `INF20\_indark' and . We are grateful to Nils Sch{\"o}neberg for very useful clarifications on \texttt{CLASS} and \texttt{CAMB} comparison. We thank the authors of \citet{Verde17}, for providing us the covariances of the EUp-15 used in this work. The authors are grateful to Vivian Poulin and Riccardo Murgia for useful comments on the draft and providing us with information from their work \citet{Poulin18}.

\bibliographystyle{apsrev4-1}
\bibliography{bibliografia}

\newpage
\appendix
\section{CMB priors}
\label{sec:Appendix}

\begin{figure}[!ht]
    \centering
    \includegraphics[scale=0.5]{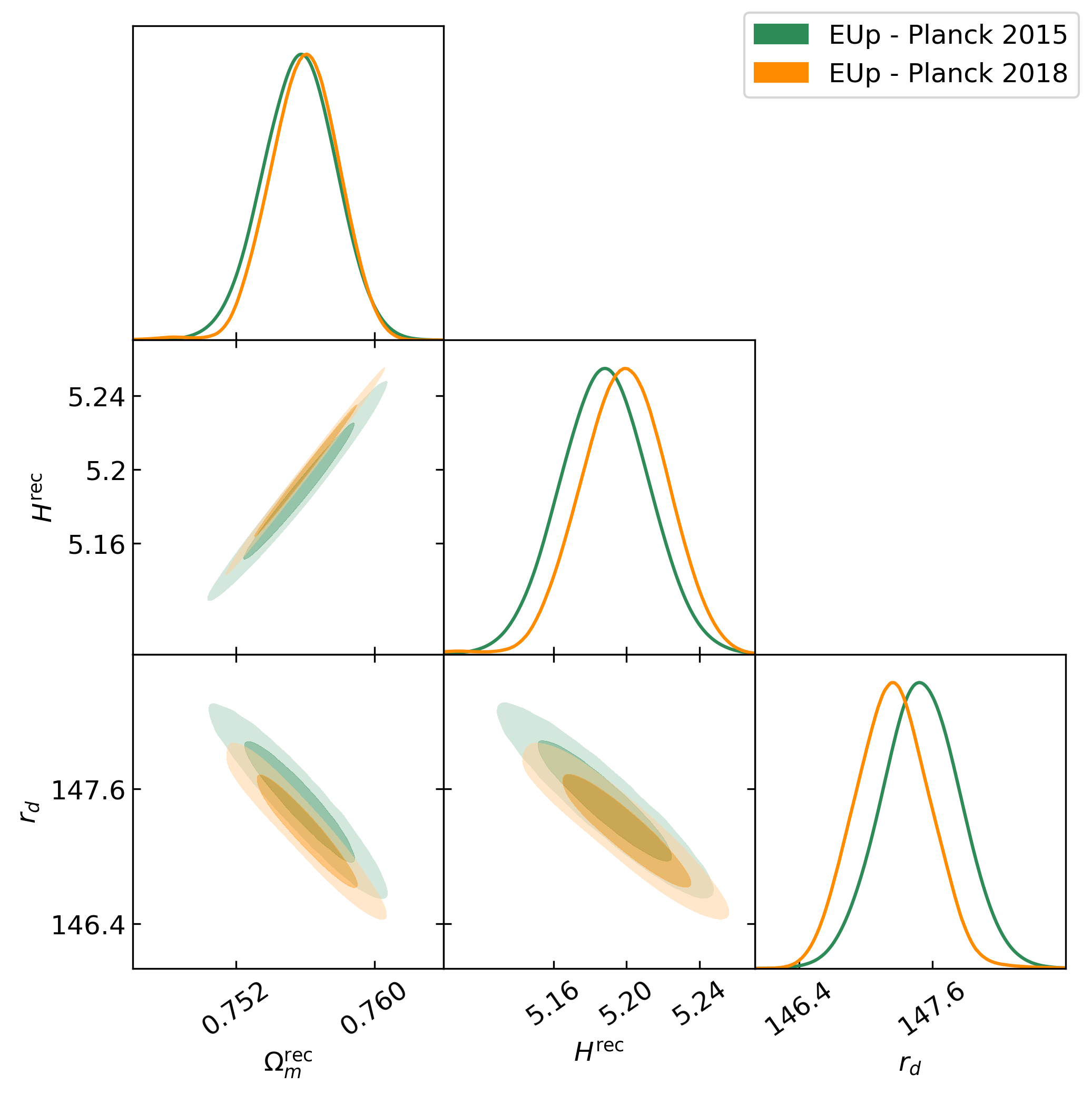}
    \caption{Comparison of the early-Universe constraints between the $\textit{Planck}$ 2015, \citet{Verde17} analysis and our analysis with $\textit{Planck}$ 2018 likelihoods. $H^{\rm rec} $ and $r_{\rm d}$ are reported in $\textrm{Mpc}^{-1}$ and $\textrm{Mpc}$, respectively.}
    \label{fig:wCDM_EUp}
\end{figure}

In this section we firstly report the covariances matrices of distance priors used in the analysis, which reproduce the joint analyses. As it has been earlier asserted in \cite{Zhai19, Chen18}, we find the distance priors sufficiently capable of replacing the complete CMB data analysis. We perform simple importance sampling\footnote{We use the \texttt{ChainConsumer} \citep{Hinton16} package to perform the importance sampling validation. The code is publicly available at \href{https://github.com/Samreay/ChainConsumer/tree/Final-Paper}{https://github.com/Samreay/ChainConsumer/tree/Final-Paper} } analysis to validate the same.

{\renewcommand{\arraystretch}{1.5}%
    \centering

\begin{table}[!ht]
   \caption{Mean values and the corresponding covariances of the distance priors for the complete CMB (TT,TE,EE+low-$l$+lowE+$lensing$) dataset. Here $z_{*} = 1089.79 \pm 0.26$.}
   \label{tab:CMB}
    \begin{tabular}{cccccc}      
        \hline
        \hline
        Observable & Mean & $\sigma_i$ &  \multicolumn{3}{c}{$r_{ij}$} \\ 
        \hline
$R(z_{*})$   & 1.7477 & 0.0041 & 1. & 0.44 & -0.65 \\
$l_A(z_{*})$   & 301.74 & 0.08  & 0.44 & 1. & -0.33 \\
$10^2 \Omega_b h^2$   & 2.243 & 0.015   & -0.65 & -0.33 & 1. \\
\hline
\hline
    \end{tabular}

\end{table}
}

{\renewcommand{\arraystretch}{1.5}%
    \centering

\begin{table}[!ht]
   \caption{Same as \cref{tab:CMB}, excluding the low-$l$+lowE datasets (i.e, CMB [high-$l$]). Here $z_{*} = 1089.69 \pm 0.31$.}
   \label{tab:CMBe}
    \begin{tabular}{cccccc}      
        \hline
        \hline
        Observable & Mean & $\sigma_i$ &  \multicolumn{3}{c}{$r_{ij}$} \\ 
        \hline
$R(z_{*})$   & 1.7418 & 0.0053 & 1. & 0.48 & -0.72 \\
$l_A(z_{*})$   & 301.68 & 0.09  & 0.48 & 1. & -0.37 \\
$10^2 \Omega_b h^2$   & 2.247 & 0.016   & -0.72 & -0.37 & 1. \\
\hline
\hline
    \end{tabular}

\end{table}
}

In \Cref{fig:wCDM_EUp}, we contrast the constraints on the early-Universe quantities as were reported in \citet{Verde17}, with the analysis replicated here using the $\textit{Planck}$ 2018 likelihoods. We find a very mild shift in the quantities mostly being consistent with the earlier results. The corresponding covariances for the $\textit{Planck}$ 2018 are reported in \Cref{tab:CMB-E18} and can be compared against \Cref{tab:CMB-E15}, presented in \Cref{sec:Data}. This small shift in the constraints is in accordance with the difference in the physical densities estimated from \textit{Planck} 2015 \cite{Planck15_DE} to \textit{Planck} 2018 \cite{Planck18_parameters}. Note that we have not validated the agreement of the EUp-18 across $\Lambda$CDM and $k\Lambda$CDM model, as reported in \citetalias{Verde17}, however expect it to remain, owing to the minimal variation in the 2015 and 2018 \textit{Planck} likelihoods.

{\renewcommand{\arraystretch}{1.5}%
    \centering

\begin{table}[!ht]
   \caption{Mean values and the corresponding covariances of the EUp-18. $H^{\rm rec} $ and $r_{\rm d}$ are reported in $\textrm{Mpc}^{-1}$ and $\textrm{Mpc}$, respectively. }
   \label{tab:CMB-E18}
    \begin{tabular}{ccccccc}      
        \hline
        \hline
        Observable & Mean & $\sigma_i$ &  \multicolumn{4}{c}{$r_{ij}$} \\ 
        \hline
$10^2 \Omega_b h^2$   & 2.246 & 1.60 $\times 10^{-2}$ & 1. & -0.73 & -0.71 & 0.39 \\
$10^2 \Omega_c h^2$   & 11.90 & 1.70 $\times 10^{-1}$ & -0.73 & 1. & 0.99 & -0.91 \\
$H^{\rm rec}$   & 5.198 & 2.40 $\times 10^{-2}$  & -0.71 & 0.99 & 1. & -0.92 \\
$r_d $   & 147.24 & 0.34 & 0.39 & -0.91 & -0.92 & 1. \\

\hline
\hline
    \end{tabular}

\end{table}
}

\begin{figure*}[!ht]
    \centering
    \includegraphics[scale=0.8]{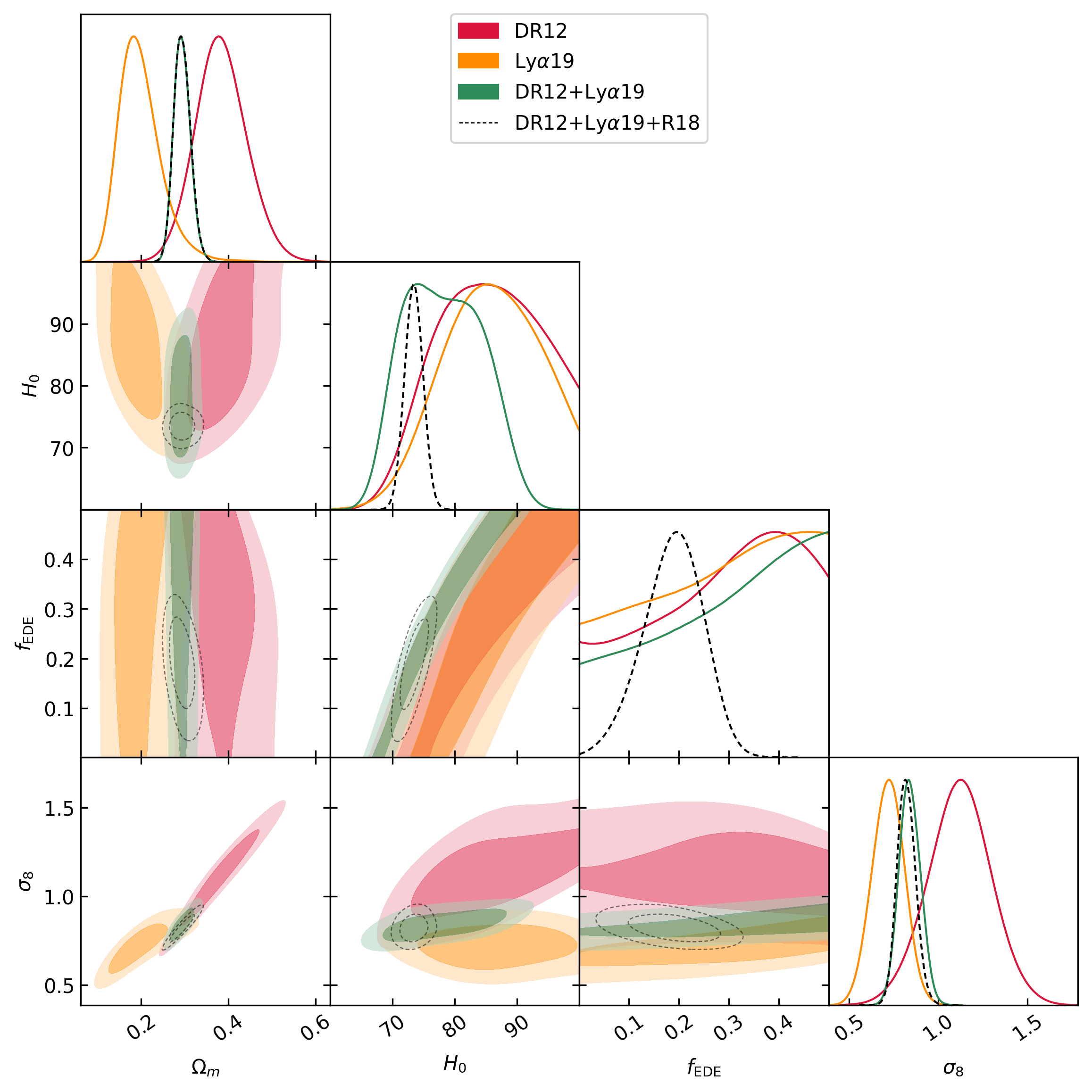}
    \caption{We present the 68\% and 95\% C.L. contours for the EDE constraints obtained form the individual BAO datasets and their joint analysis. }
    \label{fig:EDE_BAO}
\end{figure*}

\section{BAO constraints on the EDE model}
\label{sec:Appendix2}
In this section we present the complete constraints for the EDE model. The analysis is performed through the 1pEDE formalism presented in \citet{Smith20}. As expected, we do not find any constraints on the $f_{\rm EDE}$ using the individual BAO datasets, and even in the joint analysis. However, also interesting to note that the parameter tends towards larger values in contrast to the lower values preferred from the CMB data. While the $\sigma_8$ parameter is not necessarily constrained by the BAO datasets, the disagreement is induced especially by  the $\Omega_m$ limits. In \Cref{fig:EDE_BAO}, we show the constraint for the EDE model using the individual datasets and the joint analysis. More importantly, with the inclusion of the \citetalias{Riess18_H0} in the analysis we find the limits of $f_{\rm EDE} =0.189^{+0.065}_{-0.054}$ (neglecting the disagreement in the BAO datasets for the moment). As we follow 1pEDE formalism of \citet{Smith20}, we can immediately see that our limits are in very good agreement with the constraints presented therein.

\end{document}